# Two-Photon Interference from an InAs Quantum Dot Emitting in the Telecom C-Band

*Jaewon Kim, Jochen Kaupp, Yorick Reum, Giora Peniakov, Johannes Michl, Felix Kohr, Monika Emmerling, Martin Kamp, Yong-Hoon Cho,\* Tobias Huber-Loyola, Sven Höfling,\* and Andreas T. Pfenning\**

**Two-photon interference from an InAs/InAlGaAs quantum dot (QD) emitting in the telecom C-band with a raw two-photon interference visibility of $V_{HOM} = (71.9 \pm 0.2)$ % is demonstrated. This is achieved by a two-fold approach: an improvement of the molecular beam epitaxial growth for better QDs, and integration of the QDs into an optical circular Bragg grating resonator for a Purcell enhancement of the radiative decay rate. The quantum optical properties of the fabricated device are studied by means of time-correlated single-photon counting under quasi-resonant excitation of the charged exciton line. A reduced lifetime of $T_1 = (257.5 \pm 0.2)$ ps is found corresponding to a Purcell factor of $F_p \geq (4.7 \pm 0.5)$. Pronounced anti-bunching of the second-order autocorrelation function at zero time delay $g^{(2)}(0) = (0.0307 \pm 0.0004)$ confirms the single-photon emission character. The two-photon interference is demonstrated with an unbalanced Mach-Zehnder interferometer in the Hong-Ou-Mandel configuration. Strategies are discussed on how to further improve the indistinguishability, and provide a survey of the state-of-the-art.**

## 1. Introduction

A source of indistinguishable single photons is an essential building block for photonic quantum information science and technologies.[1–3] Highly indistinguishable photons are required

J. Kim, Y.-H. Cho
Department of Physics
Korea Advanced Institute of Science and Technology (KAIST)
Daejeon 34141, Republic of Korea
E-mail: yonghcho@kaist.ac.kr

J. Kim, J. Kaupp, Y. Reum, G. Peniakov, J. Michl, F. Kohr, M. Emmerling, M. Kamp, T. Huber-Loyola, S. Höfling, A. T. Pfenning
Physikalisches Institut
Lehrstuhl für Technische Physik
Julius-Maximilians-Universität Würzburg
D-97074 Würzburg, Germany
E-mail: sven.hoefling@uni-wuerzburg.de; andreas.pfenning@uni-wuerzburg.de

to achieve high-fidelity two-photon interference, first described by Hong, Ou, and Mandel.[4] A whole toolbox of quantum information processing schemes in quantum optics is based on two-photon interference, such as Bell state measurement,[5] quantum teleportation,[6,7] and entanglement swapping.[8] Furthermore, two-photon interference can be used for the generation of higher photon-number entangled states, such as the biphoton N00N state[9] or heralded three-photon Greenberger-Horne-Zeilinger (GHZ)[10,11] states. Also, two-photon interference-based fusion of resource states can generate photon cluster states of large photon number and higher dimension,[12–15] which is crucial for measurement-based quantum computing.[16]

In this regard, epitaxially grown semiconductor quantum dots (QDs) have become a launchpad for quantum photonic applications.[17] In particular, InAs/GaAs QDs typically emitting in the wavelength range of 900–960 nm have been demonstrated as near-ideal sources of indistinguishable single photons.[18–20] To achieve the emission of indistinguishable single photons from these sources, it is advantageous to integrate the QDs into a photonic resonator, such as micropillars,[21–23] open cavities,[24] photonic crystal cavities[25] and circular Bragg gratings (CBG).[26–28] A quantum dot resonant with such a microcavity in the weak coupling regime will experience a reduced excitonic lifetime via the Purcell effect.[29] As a result, the indistinguishability of the emitted photons is enhanced as the excitonic lifetime becomes shorter compared to the pure-dephasing time.[30]

For fiber-based quantum photonic applications, emission in the telecom C-band (1530–1565 nm), where the loss in standard optical fibers is at a minimum, is crucial. In addition, this wavelength is well compatible with low-loss integrated quantum circuit platforms like silicon, SiN, or lithium niobate. However, the photon indistinguishability from such C-band emitting QD sources has not yet achieved the degree of InAs/GaAs QDs. Current state-of-the-art indistinguishability of C-band emitting QDs is obtained from InAs QDs grown on a metamorphic InGaAs layer and InAs/InP QDs. An indistinguishability of up to $V_{HOM} = 34.9\%$ is reported for InAs QDs on a metamorphic buffer, integrated within CBG.[31] A similarly high indistinguishability of 35% is reported from InAs/InP QDs.[32]

Here, we tackle this with a two-fold strategy, based on a well-calibrated and improved semiconductor growth exploiting ternary digital alloying of the bulk semiconductor material and Stranski-Krastanov growth of the QDs,[33] as well as the QD integration into CBG resonators with high Purcell enhancement.[34] Emission in the telecom C-band is observed with a charged exciton line at $\lambda$ = 1544.5 nm as a dominant transition. We investigate the multi-photon probability of the emission by the second-order temporal correlation function obtained from a Hanbury Brown and Twiss (HBT) experiment. The degree of photon indistinguishability is probed by two-photon interference in the Hong-Ou-Mandel (HOM) configuration.

## 2. Sample Design and Fabrication

The nanophotonic structure is designed by using finite-difference time-domain simulations to identify the ideal resonator geometry. Subsequently, the QD sample is grown by gas-source molecular beam epitaxy, and the circular Bragg grating resonators are defined by electron beam lithography and dry-chemical etching. For a detailed description of the simulation as well as the growth and fabrication, please refer to Ref. [34] Two modifications to the growth are made.

First, the $In_{0.53}Al_{0.23}Ga_{0.24}As$ barrier layer, which confines the QDs, is grown by ternary digital alloying. That is, the quaternary is comprised of alternating monolayers of the ternary compounds $In_{0.52}Al_{0.48}As$ and $In_{0.53}Ga_{0.47}As$. This growth technique results in an improved homogeneity and reduced clustering of the barrier material.[35] Furthermore, the composition of the termination layer, which comes prior to the InAs QD growth, can be controlled precisely. In the present sample, $In_{0.53}Ga_{0.47}As$ is chosen as the termination layer as it provides the highest surface mobility of InAs.

Second, precise adjustment of the InAs thickness used for QD growth allows for overall lower QD density and higher quality. The desired InAs thickness is identified by first growing a calibration sample with an InAs gradient across the wafer. This gradient is achieved by stopping the wafer rotation during the growth of the InAs layer, resulting in an inhomogeneous deposition profile from the indium effusion cell. The exact desired InAs thickness is extracted from the position with desired QD properties on the calibration sample and implemented in the actual sample with rotation. The rotation frequency is matched with the InAs deposition time to ensure a homogeneous distribution throughout the wafer. Growth rate fluctuation of the effusion cell between days does carry the need for a small InAs growth series to achieve the desired InAs thickness on the full wafer.

The increased surface migration, together with a precise control of the InAs thickness enables a decrease in ripening time from 150 to 30 s where QD growth in the telecom C-band can be achieved. The reduced ripening time potentially leads to less incorporation of unwanted contaminants from the residual background contamination in the growth chamber. Morphology measurements on a calibration sample with InAs gradient, by means of atomic force microscopy, show a density of the larger QDs, expected to emit in the telecom C-band in the range of $\approx 10^{-8}$ $cm^2$ with reduced density of smaller structures compared to Ref. [34] This is also evident in a reduced overall spectral density in micro photoluminescence between the desired telecom C-band and the wetting layer emission $\approx$1200 nm.

The nanofabrication is also described in Ref. [34] with the modification that 490 nm of sputter-deposited $AlO_X$ is used as a dielectric layer instead of $SiO_2$ as it shows more reliable adhesion between the layers resulting in higher yield during the membrane fabrication process.

## 3. Optical Device Properties

### 3.1. Photoluminescence Characteristics of QD CBG Device

The initial optical device characterization is performed by means of micro-photoluminescence (μPL) spectroscopy under a nonresonant excitation condition. The sample is mounted in a liquid helium flow cryostat at a base temperature of $T$ = 4.3 K. A fiber-coupled continuous-wave (cw) laser is used for above-band excitation at a wavelength of $\lambda_{exc}$ = 660 nm.

The μPL spectrum of the device is presented in **Figure 1a**. Under low excitation power of $P$ = 150 nW, individual transition lines in the telecom C-band range signify recombination from different excitonic complexes in a single QD. To measure the spectrum of the cavity mode of the CBG resonator, a higher excitation power of $P$ = 100 μW is used to saturate emitters, thereby illuminating the entire cavity mode. From this measurement, the center wavelength of the cavity mode is determined to be $\lambda_{cav}$ = 1545.5 nm, and the full width at half maximum (FWHM) of the mode is $\Delta\lambda_{FWHM,cav}$ = 6.0 nm. Due to a small unwanted ellipticity of the CBG, an energy splitting of the two orthogonally linear polarized fundamental modes is resolved. One of the modes, which is defined as horizontally polarized (H) mode, shows a center wavelength of $\lambda_{cav,H}$ = 1544.2 nm and a width of $\Delta\lambda_{FWHM,cav,H}$ = 4.3 nm. The other mode, which is defined as vertically polarized (V) mode, shows a center wavelength of $\lambda_{cav,V}$ = 1546.7 nm and a width of $\Delta\lambda_{FWHM,cav,V}$ = 3.8 nm. The dominant transition line and H-mode have a small detuning of 0.3 nm, which is a good spectral alignment, given the width of the cavity mode.

To determine the origin of the transition lines, we perform excitation power- and detection polarization-dependent μPL measurements. For the polarization series, the excitation power is set to $P$ = 1 μW. Based on these measurements, the dominant emission line at 1544.5 nm is identified as the charged exciton transition ($X^{+/-}$), while lines at 1541.0 and 1547.6 nm are attributed to the neutral exciton (X) and biexciton transitions (XX) of the single QD, respectively. As shown in the polarization series in Figure 1b, oscillatory features are observed for the X and XX transition lines, due to the exciton fine-structure splitting $S$ of (22.9 ± 0.1) μeV, whereas the other lines do not exhibit such behavior. Our analysis is focused on the dominant charged exciton transition line. A degree of linear polarization of (0.495 ± 0.006) from the dominant charged exciton line is observed, which is attributed to a probable spatial displacement from the cavity center[36] and the ellipticity of the cavity.[37]

### 3.2. Quasi-Resonant Excitation and Purcell Enhancement

We perform a photoluminescence excitation (PLE) measurement to determine higher energy resonances of the QD. For this, an optical parametric oscillator (OPO), pumped by a Ti: sapphire pico-second pulsed laser with a 76 MHz repetition rate is used to excite the QD with the desired wavelength. During the PLE measurements, the emission spectrum is recorded while scanning the excitation wavelength. A clear resonance, presumed to be the p-shell resonance of QD, is observed at $\lambda_{exc} = 1406$ nm [See **Figure** 2a].

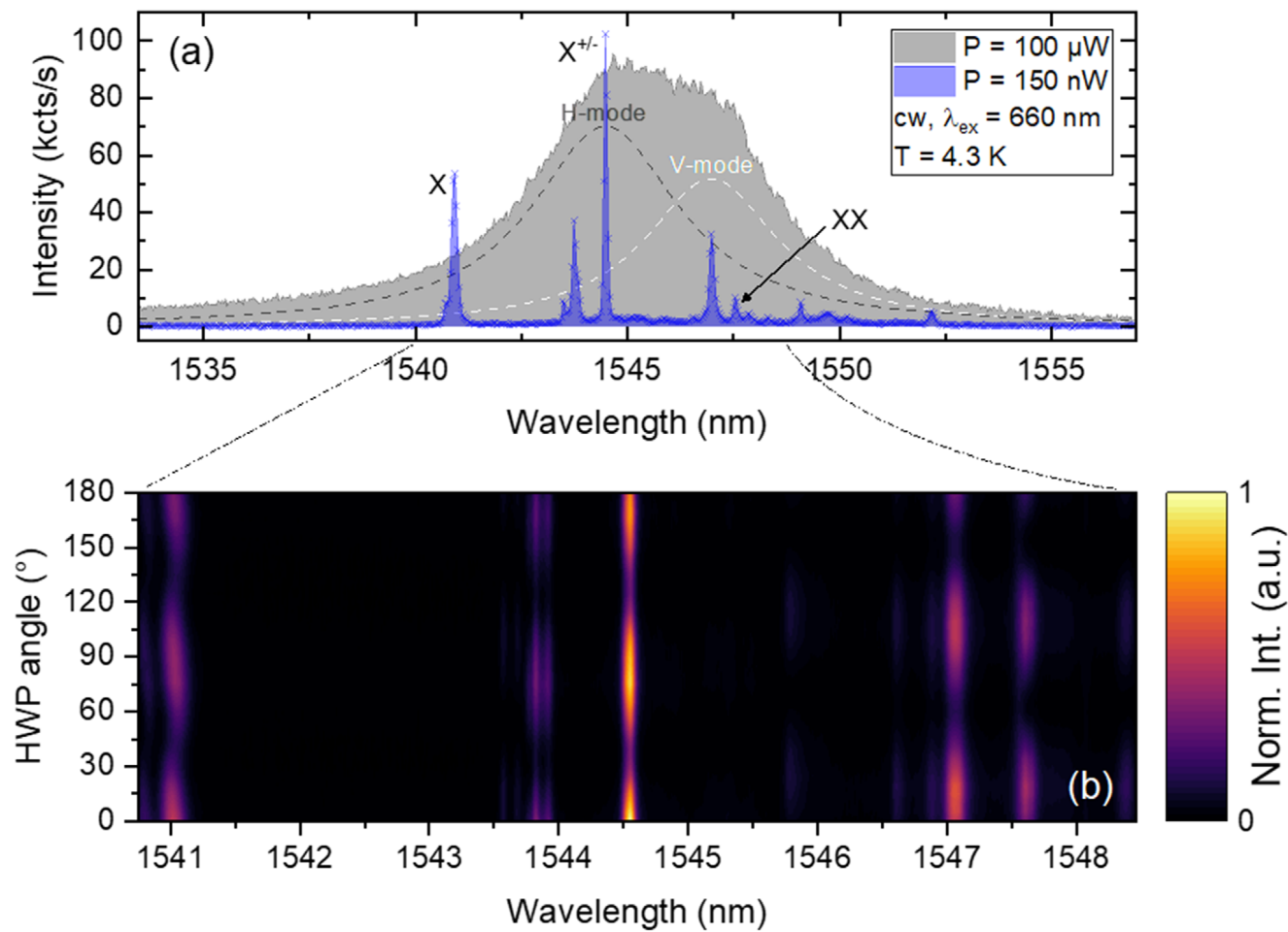

**Figure 1.** µPL characteristic of the QD-CBG device. a) µPL spectrum of the device under the above-band cw excitation with the excitation wavelength of $\lambda_{exc} = 660$ nm. High (**P** = 100 µW) and low (**P** = 150 nW) power excitations are utilized to obtain spectra of the cavity mode (gray) and the single QD emission (blue), respectively. The cavity mode spectrum is fitted by two orthogonally polarized modes (dashed lines), where the center wavelength and FWHM of each polarized mode are extracted from polarization-dependent µPL measurement at high power (**P** = 80 µW) excitation. The dominant charged exciton line and the H-mode are spectrally well aligned with 0.3 nm detuning. b) Color plot of polarization-dependent µPL under low power (**P** = 1 µW) excitation. The dominant line at 1544.5 nm is identified as a charged exciton ($X^{+/-}$) transition. The lines at 1541.0 and 1547.6 nm are attributed to transitions of a neutral exciton (X) and a biexciton (XX).

Figure 2b shows the µPL spectrum under quasi-resonant excitation ($\lambda_{exc}= 1406$ nm) recorded at an incident excitation power of $P = 550$ nW. Under these excitation conditions, the charged exciton transition line becomes dominant over other transitions, with a FWHM linewidth of $\Delta E_{FWHM} = 33.8$ µeV directly determined from the data. The Voigt fit yielded a Lorentzian width of $\Delta E_L = (12 \pm 1)$ µeV and a Gaussian width of $\Delta E_G = (26.2 \pm 0.9)$ µeV, where $(18.0 \pm 0.5)$ µeV is the measured spectrometer resolution. The error of the linewidth is given from the standard error of the fit. Compared to above-bandgap excitation, background emission is suppressed. This narrow linewidth is an improvement of almost one order of magnitude compared to our previous work,[34] and is on par with the best reported values for telecom C-band QDs.[31,38–40]

Ref. [41] has shown that coherent scattering can produce sub-linewidth photons in the Telecom-C band, even in the inelastic scattering regime. However, since the coherence in the inelastic scattering regime is power dependent,[42] we refrain from comparing our linewidth with the best linewidth measured incoherent scattering. At saturation, the coherence reported in coherent scattering is comparable with what we report.

The emission of the charged exciton is coupled to an optical fiber and spectrally isolated with a tunable narrow bandpass filter. The bandwidth of the flat-top band pass filter is set to $\Delta E = (67 \pm 1)$ µeV, which is more than double the linewidth of the charged exciton emission. This is measured by filtering a broadband signal analyzing its spectrum and fitting a convolution of the expected flat top with the measured spectrometer resolution. The error is estimated from the standard error of the fit-ting parameters. This filtered emission is detected by superconducting nanowire single-photon detectors (SNSPDs). The temporal FWHM of the instrument response function is measured as 38 and 36 ps for SNSPD channels 1 and 2, respectively.

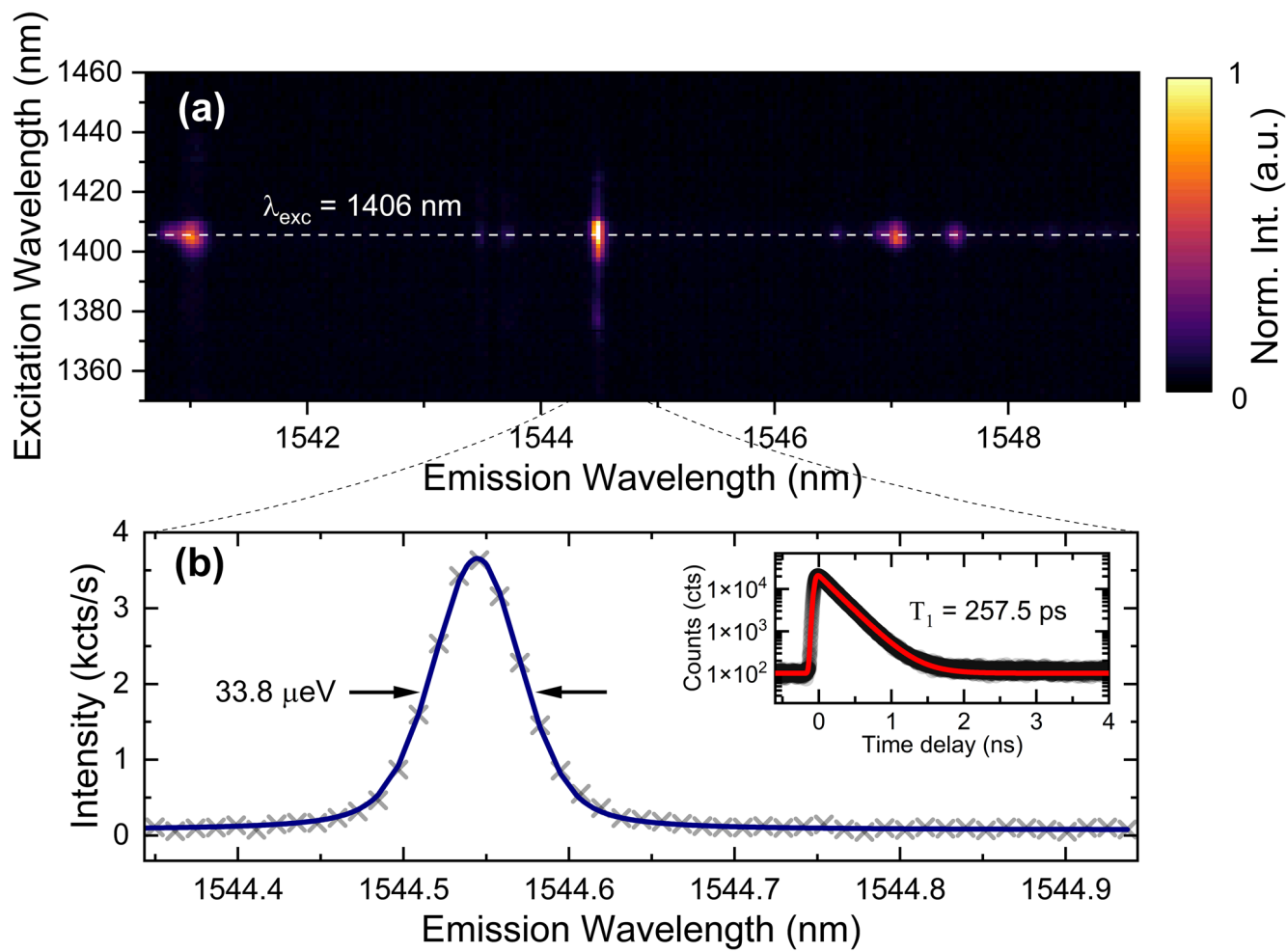

**Figure 2.** PLE scan and spectrum under quasi-resonant excitation. a) PLE color plot obtained by changing the excitation laser wavelength with an OPO down conversion. Clear resonance is observed at $\lambda_{exc}$ = 1406 nm, which corresponds to an energy difference of $E_{exc} - E(X_0)$ = 77 meV from the exciton transition. We presume this resonance is the p-shell transition of the QD. b) Spectrum under quasi-resonant excitation with an excitation power of $P$ = 550 nW. The dominant charged exciton transition shows a linewidth of $\Delta E_{FWHM}$ = 33.8 μeV. The solid line illustrates the Voigt fit with Lorentzian width $\Delta E_L$ = (12 ± 1) μeV of and Gaussian width of $\Delta E_G$ = (26.2 ± 0.9) μeV. The inset shows the excitonic decay histogram under quasi-resonant excitation. The histogram shows a mono-exponential decay with an excitonic lifetime of $T_1$ = (257.5 ± 0.2) ps. Compared with the previously studied bare excitonic lifetime in a homogenous medium from Ref. [34] the excitonic lifetime of the QD-CBG device is reduced by a Purcell enhancement factor of $F_P \geq$ (4.7 ± 0.5).

The lifetime of the charged exciton line is measured utilizing time-correlated single-photon counting (TCSPC). The excitonic decay of the charged exciton is obtained by measuring the time delay between the laser trigger and the photon detection event at the SNSPD channel 1 [see Figure 2b inset]. By fitting the decay histogram, the excitonic lifetime of the charged exciton line is determined to be $T_1$ = (257.5 ± 0.2) ps. By comparing the reference lifetime of $T_1$ = (1210 ± 115) ps for QDs in a planar membrane from our previous study[34] to the reduced excitonic lifetime, a Purcell enhancement factor of $F_P \geq$ (4.7 ± 0.5) is obtained. Please note that this is a conservative estimate of the Purcell factor as reported values of the excitonic lifetime in comparable QDs are typically at about $T_1 \approx$ 1.5 − 2.0 ns,[43–45] which agrees with the calculated value of $T_1$ = 1.7 ns we obtained from numerically simulating the 3D electron and hole wavefunction.[46]

## 4. Quantum Optical Properties of the QD-CBG Device

### 4.1. Multi-Photon Probability

The multi-photon probability of the charged exciton line is obtained by measuring the second-order autocorrelation function in HBT configuration and acquiring the coincidence histogram as a function of the time delay between the photon clicks from different beam splitter output ports. The device is excited under a quasi-resonant condition with $\lambda_{exc}$ = 1406 nm and with an excitation power of $P$ = 150 nW, which corresponds to a working point of ≈2/3 of saturation counts, it is impor-tant to mention that the multi-photon probability and the indistinguishability are in general not independent of the excitation laser power. Two SNSPD channels record count rates of 8.5 and 8.3 kcts s$^{-1}$, respectively (with an integration time of 72 min).

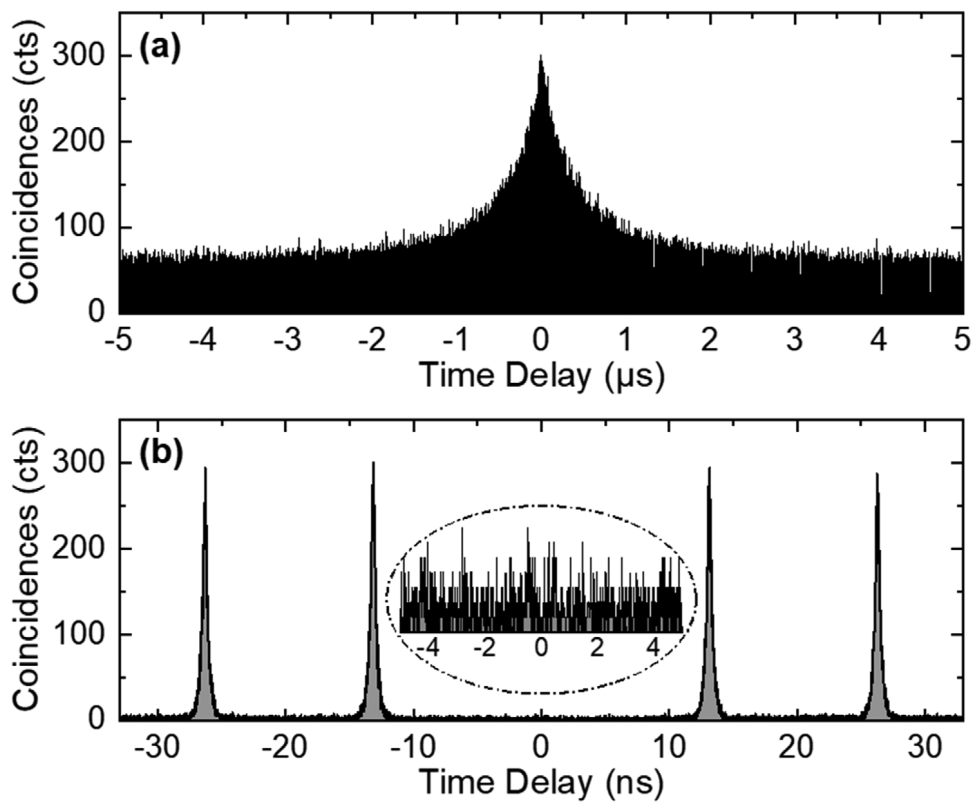

**Figure 3.** Second-order autocorrelation function of the charged exciton emission line measured in HBT configuration. a) Coincidence histogram as a function of time delay across a 10 μs temporal window from -5 μs to +5 μs. A considerable blinking effect with a blinking time of $T_{blink} = (506 \pm 3)$ ns and blinking strength of $A_{blink} = (3.70 \pm 0.02)$ is observed. b) Coincidence histogram of the center peak and the first and second side peaks. The suppressed coincidences of the center peak indicate a low multi-photon probability of $g^{(2)}_{HBT,\,3\,ns}(0) = (0.0307 \pm 0.0004)$. The inset shows the coincidence histogram of the center peak.

In **Figure 3**a, the coincidence histogram of the HBT measurement is shown across a broad window from -5 μs up to 5 μs of time delay. Pronounced blinking is observed. To describe the blinking behavior, each coincidence peak is integrated except the center peak. Integrated peak areas are fitted by a model suggested in Ref. [47] given as $h_{m\neq 0} = h_0 (1 + A_{blink}\, e^{-|\tau|/T_{blink}})$, where $h_{m\neq 0}$ is the $m^{th}$ coincidence peak area and $h_0$ is the peak area at $|\tau| \gg T_{blink}$. A blinking time of $T_{blink} = (506 \pm 3)$ ns and a blinking strength of $A_{blink} = (3.70 \pm 0.02)$ are obtained from the fit. This blinking behavior can be due to several reasons such as charge transferring to defect states, which could change the total charge of the quantum dot; spin-flip to a dark state; excitation laser wavelength fluctuation; and Auger effect.[48,49] From the blinking strength, we can estimate the upper bound of the quantum efficiency (probability of photon emission from the spectrally filtered state upon each excitation pulse) to $\eta_{blink} \leq \frac{1}{1+A_{blink}} = (21.3 \pm 0.1)\%$.[32,47]

Figure 3b shows the HBT coincidence histogram on a zoomed-in time scale to observe the g$^{(2)}$(0) peak, spanning across the two nearest side peaks for positive and negative time delay, respectively. Pronounced antibunching is observed as there is no coincidence peak at zero time delay. This antibunching behavior of the center peak indicates that the charged exciton emission has a low multi-photon probability. A raw multi-photon probability of $g^{(2)}_{HBT,\,3\,ns}(0) = (0.0307 \pm 0.0004)$ is obtained from a 3 ns coincidence window, comparing the zero peak to the first side peaks. This is a conservative estimate for a blinking-free $g^{(2)}(0)$ value, which reflects the multi-photon component of the source as compared to a source with the same excitation probability and without blinking. A 3 ns coincidence window is sufficient to fully contain all the counts of the coincidence peak because of its short lifetime of $T_1 = (257.5 \pm 0.2)$ ps. For applications where the multi-photon probability is compared with the average time integrated photon number (e.g., quantum key distribution), we cannot compare with a blinking-free source, but a normalization based on the Poissonian level is more conclusive. For those cases $g^{(2)}(0) = (0.136 \pm 0.002)$. The error in both cases is estimated based on the Poissonian error of the coincidence counts without fit or background subtraction.

### 4.2. Two-Photon Interference

To measure the two-photon interference visibility of subsequently emitted photons, the 50:50 beamsplitter that comprises the HBT setup is exchanged for an unbalanced Mach-Zehnder interferometer to measure HOM interference. After splitting the photons 50:50 using a Wollaston prism, both arms of the interferometer contain a half-wave plate and a quarter-wave plate for polarization control between the arms. To calibrate the polarization, a cw laser with the same wavelength as the charged exciton emission is sent to the interferometer and measured by a polarimeter. The photons from both arms are coupled into optical fibers. Only one interferometer arm has a delay line, causing a relative time delay of 13.16 ns, which is the period of the photon train. This delay makes two subsequently emitted photons indistinguishable in time and thereby allows one photon to interfere with another photon from the next pulsed excitation. At the end of the interferometer, a 50:50 fiber beam splitter is utilized.

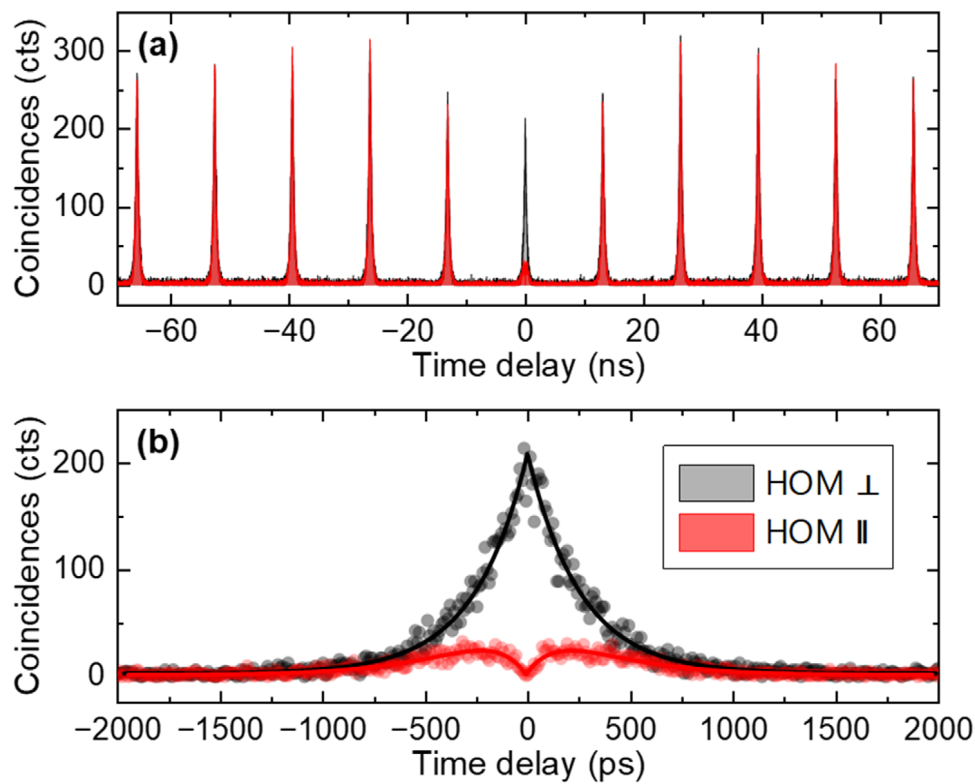

**Figure 4.** Two-photon interference coincidence histogram of consecutive photons from the charged exciton in a co- (red) and cross- (black) polarization configuration. a) Coincidence histogram of center peak and 10 side peaks. The side peaks match well between the co- and cross-polarized configuration. b) Center peak histogram. Suppression of the center peak in the co-polarization configuration compared to the cross-polarization configuration indicates the HOM effect of consecutive photons. Scattered data points are raw data (cross-polarized histogram is normalized). Solid lines are from the fitting for co- and cross-polarized coincidence histograms. We obtain $V_{HOM, 3\,ns} = (71.9 \pm 0.2)\%$, which is the current state-of-the-art for telecom C-band QD single photon sources. By fitting the co-polarized coincidence histogram, a coherence time of $T_2 = (381.4 \pm 21.8)$ ps is obtained.

In the experiment, the QD-CBG device is excited under the same excitation and collection conditions as for the HBT measurement. The count rates for the two SNSPD channels are measured to be 6 kcts s$^{-1}$ for the co-polarized configuration (with an integration time of 158 min) and 4 kcts s$^{-1}$ for the cross-polarized configuration (with an integration time of 323 min). Photons from two outputs of the interferometer are detected at the SNSPDs, and a coincidence histogram as a function of detection time delay is obtained. In **Figure 4**, the obtained coincidence histograms for co- and cross-polarized are shown. Since orthog-onally polarized photons are maximally distinguishable, two-photon interference visibility for pulsed excitation is defined by comparing the co-polarized coincidence with the cross-polarized coincidence area of the center peak as $V_{HOM} = 1 - \frac{A_\parallel}{A_\perp}$. For a blinking-free source, the center peak of cross-polarization should be 1/2 of the Poissonian level. In our case, due to the blinking behavior, the center peak contains more than 1/2 of the coincidence counts of the second side peaks. The first side peaks are expected to be at 75% of the Poissonian level due to the introduced Mach-Zehnder interferometer with a delay corresponding to exactly one laser repetition rate. All the peaks afterward are from uncorrelated events, as in the HBT measurement. To ensure that the co- and cross-polarized measurements are comparable, we normalize them to the peaks from uncorrelated events. Since we compare directly the events from co- and cross-polarized measurements, the blinking behavior is irrelevant to our data evaluation.

The reduction of the coincidence near zero-time delay for the co-polarized case clearly demonstrates the two-photon interference of the charged exciton emission. The two-photon interference visibility for a 3 ns integration window is $V_{HOM, 3\,ns} = (71.9 \pm 0.2)$ %. The uncertainty is estimated as the Poisson standard deviation of the coincidence counts. It is noteworthy that, to the best of our knowledge, this is the highest reported two-photon interference visibility on any QD device

Table 1. Survey on the reported HOM visibility, lifetime, and linewidth of QD single-photon sources emitting in the Telecom C-band.

| QD Type | Growth | Device | Excitation | $V_{\text{HOM}}$ | $T_1$ | $\Delta E_{\text{FWHM}}$ | Refs. |
|---|---|---|---|---|---|---|---|
| InAs/InAlGaAs | MBE | CBG | QR (p-shell) | $V_{\text{3 ns}} = 71.9\%$ | 257.5 ps | 28.9 µeV | This work |
| InAs/InP | MOVPE | Mesa | TPE | $V_{\text{4 ns}} = 35\%$ | 340 ps | 47 µeV | [32] |
| InAs/MMB | MOVPE | CBG | LA, SUPER | $V = 34.9\%$ (LA) | 500 ps | 20.0 µeV | [31] |
| | | | | $V = 10.4\%$ (SUPER) | | | |
| InAs/InP | MOVPE | CBG | LO | $V_{\text{fit}} = 19.3\%$ | 530 ps | 87 µeV | [45] |
| InAs/MMB | MOVPE | CBG | p-shell | $V_{\text{fit}} = 8.10\%$ | 520 ps | 65 µeV | [44] |
| InAs/MMB | MOVPE | Planar | RF | $V_{\text{fit}} = 14.4\%$ | 1010 ps | ≈30 µeV | [52] |

Subscript of the visibility indicates the integration window size used in each study. MMB: metamorphic buffer, MBE: molecular beam epitaxy, MOVPE: metal-organic vapor phase epitaxy, CBG: circular Bragg grating, QR: quasi-resonant excitation, TPE: two-photon excitation, LA: LA-phonon assisted excitation, SUPER: swing-up population quantum emitter population, LO: LO-phonon assisted excitation, RF: resonant fluorescence.

emitting in the telecom C-band. Recent developments in enhancing the indistinguishability of telecom C-band QD devices are described in **Table 1**. Our QD-CBG device shows a small broadening combined with a short lifetime, compared to recent works on telecom C-band QDs. The high indistinguishability results from both the reduced linewidth and the shortened lifetime. The measured linewidth suggests lower indistinguishability than we measured. This discrepancy arises from the different timescales of the two measurements. The linewidth measurement includes all dephasing mechanisms within the measurement time of 10 s, while the HOM measurement only senses dephasing between two successively emitted photons, which is 13 ns in our case. Note that the quasi-resonant p-shell excitation is not a fully optimized excitation scheme, since p-shell to s-shell relaxation timing jitter degrades the indistinguishability.[30,50,51] Therefore, there is still room to enhance indistinguishability by using a more tailored excitation scheme.

From the co-polarized coincidence histogram, the coherence time $T_2$ as well as the pure dephasing time $T_2^*$ can be extracted. To fit the co-polarized HOM coincidence of the center peak, we use $C_{\text{HOM},\parallel}(\tau) = A(\exp(-\frac{|\tau|}{T_1}) - V\exp(-\frac{2|\tau|}{T_2}))$ as a fitting function, where $\tau$ is the time delay between the detection of the two outputs, $V$ is the maximum visibility exactly at zero time delay (1 in our case), and $T_1$ and $T_2$ are the lifetime and the photon coherence time, respectively.[53] The fitting parameter $T_1$ is fixed to the value of $T_1 = 257.5$ ps obtained from the lifetime measurement. The errors below are estimated from the standard error of the fitting parameters. From the fitting, the coherence time of $T_2 = (381 \pm 22)$ ps is derived. Based on the obtained value for the coherence time and the relation $\frac{1}{T_2} = \frac{1}{2T_1} + \frac{1}{T_2^*}$, the pure dephasing time is calculated as $T_2^* = (1470 \pm 320)$ ps. The term $\frac{1}{2T_1}$ describes the coherence from the finite lifetime of the emitter, also referred to as the Fourier limit. If this was the only factor, the HOM would show perfect interference over the whole wave packet of the photon. The additional term, the pure dephasing $T_2^*$, comes from both homogeneous broadening by coupling to the phonon bath and inhomogeneous broadening from electric or magnetic environment fluctuations. By comparing the fitted area of the co- and cross-polarized coincidences (excluding the baseline from the fitting), we obtain a two-photon interference visibility of $V_{\text{HOM,fit}} = (74.1 \pm 6.2)$ %. The two-photon interference visibility derived from this fit is slightly higher than from integrating over the 3 ns window, due to the correction for background contributions, which comes from the detection system. We forgo correcting the HOM visibility based on the $g^{(2)}(0)$ value, which is often done,[19,24] because we consider the $g^{(2)}(0)$ to be an intrinsic characteristic of the source that cannot simply be disregarded.

The deduced values for $T_1$ and $T_2$ give an insight into how the two-photon interference visibility can be further improved [see **Figure 5**]. $T_2^*$ is determined by various broadening mechanisms of the excitonic QD state, which depends on the material quality and the applied excitation scheme. In contrast, $T_1$ can be reduced by Purcell enhancement, independently from $T_2^*$. High Purcell enhancement with $F_P > 25$ is reported by deterministically fabricating a CBG on pre-measured QD in the wavelength range of 920–940 nm.[27] Assuming the $T_2^*$ is independent of the reduction of $T_1$, with the measured $T_2^*$ from our sample, we can obtain two-photon interference visibility up to $V_{\text{HOM}} > 94\%$ when the lifetime is reduced to $T_1 < 48$ ps. The corresponding Purcell factor of $F_P > 25$ is in principle also achievable with our current structure, if the QD and the resonator are spatially and spectrally well aligned. Optimized designs showing this enhancement from FDTD simulations are presented in our earlier work in Ref. [34].

## 5. Conclusion

We have demonstrated a two-photon interference visibility of $V_{\text{HOM, 3 ns}} = (71.9 \pm 0.2)$ % from the charged exciton transition of an InAs/InAlGaAs QD emitting in the telecom C-band, which is the unprecedented visibility for any QD single photon source emitting in the telecom C-band. We deduced a pure-dephasing time of $T_2^* = (1470 \pm 320)$ ps, which is ≈5.7 times as long as the Purcell-enhanced excitonic lifetime of $T_1 = (257.5 \pm 0.2)$ ps. Optimization of the InAs layer thickness and a proper termination layer of digital alloys reduce the ripening time during QD growth. This increases homogeneity and diminishes defects, eventually reducing the non-radiative broadening mechanism. We attribute these two effects, an improved dephasing and reduced lifetime, as the main reasons for obtaining high HOM visibility. This highlights that the achieved two-photon interference visibility in this work sets a milestone toward an ideal QD indistinguishable single photon source emitting in the telecom C-band.

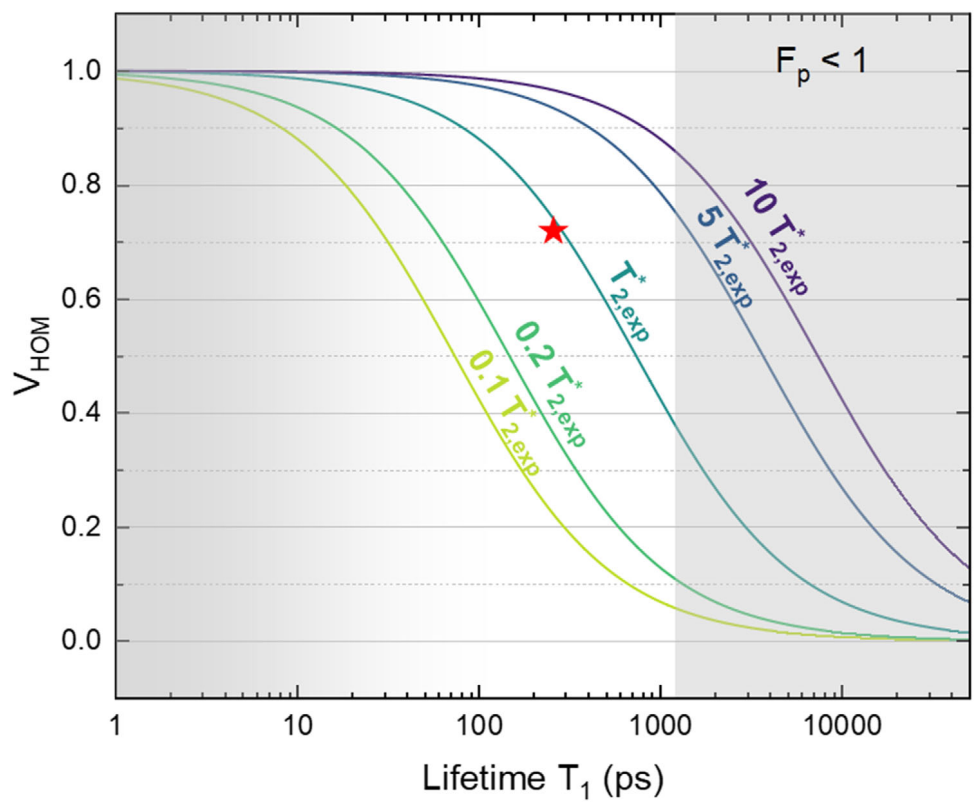

**Figure 5.** Calculated two-photon interference visibility with respect to $T_1$ and $T_2^*$. Colored lines indicate different $T_2^*$ scaled from experimentally obtained pure dephasing time of $T_{2,exp}^* = 1470$ ps. The red star represents our lifetime $T_1 = 257.5$ ps and the raw two-photon interference visibility $V_{HOM,\,3\,ns} = (71.9 \pm 0.2)$ % from this work.

## Acknowledgements

The authors acknowledge the support of the state of Bavaria and the German Ministry for Research and Education (BMBF) within Project PhotonQ (FKZ: 13N15759) and QuNET+ICLink (16KIS1975). Tobias Huber-Loyola acknowledges financial support from the BMBF within the Project Qecs (FKZ: 13N16272). The authors are furthermore grateful for the support by the National Research Foundation (NRF) of South Korea within the international joint research program on "Chip-scale Scalable Quantum Light Sources and Photonic Integrated Circuit Technology Development" (RS-2023-00284018). The authors also acknowledge the support from the free state of Bavaria, and the German research foundation (DFG) under the project Nr. INST 93/1007-1 LAGG.

## Conflict of Interest

The authors declare no conflict of interest.

## Author Contributions

J.W.K., J.K., and Y.R. contributed equally to this work. J.W.K., J.K., and Y.R. performed formal analysis, investigation, resources, and visualization and wrote, reviewed, and edited the original draft. G.P. and J.M. performed an investigation and wrote, reviewed, and edited the original draft. F.K. and M.E. performed resources, and wrote, reviewed, and edited the original draft. M.K. performed Funding acquisition and resources, and wrote, reviewed, and edited the original draft. Y.-H.C. performed funding acquisition, project administration, supervision, and validation, and wrote, reviewed, and edited the original draft. T.H.-L. performed conceptualization, formal analysis, funding acquisition, methodology, project administration, supervision, and validation, and wrote, reviewed, and edited the original draft. S.H. performed conceptualization, funding acquisition, methodology, project administration, supervision, and validation, and wrote, reviewed, and edited the original draft. A.P. performed conceptualization, formal analysis, investigation, methodology, project administration, supervision, validation, and visualization, and wrote, reviewed, and edited the original draft.

## Data Availability Statement

The data that support the findings of this study are available from the corresponding author upon reasonable request.